\documentclass[prb,superscriptaddress,onecolumn]{revtex4}

\usepackage{physics}
\usepackage{graphicx}
\usepackage{dcolumn}
\usepackage{bm}
\usepackage{siunitx}
\usepackage{booktabs}
\usepackage{amssymb}
\usepackage{amsfonts}
\usepackage[english]{babel}
\usepackage{hyperref}
\usepackage{bigints}
\usepackage{eqnarray,amsmath}
\usepackage{textcomp}

\usepackage{amsmath,amssymb}
\usepackage{lipsum}
\usepackage[usenames, dvipsnames]{color}
\usepackage{epstopdf}
 \usepackage{setspace}
\begin{document}

\title
  { Improving N\'{e}el domain walls dynamics and skyrmion stability using  He ion irradiation  }

\author{Cristina Balan}
\affiliation{Univ. Grenoble Alpes, CNRS, Institut N\'eel, 38042 Grenoble, France}
\author{Johannes W. van der Jagt}
\affiliation{Spin-Ion Technologies; 10 Boulevard Thomas Gobert, 91120 Palaiseau, France  }
\affiliation{Centre de Nanosciences et de Nanotechnologies, CNRS, Université Paris-Saclay, 10 boulevard Thomas Gobert, 91120 Palaiseau, France}
\author{Aymen Fassatoui}
\affiliation{Univ. Grenoble Alpes, CNRS, Institut N\'eel, 38042 Grenoble, France}
\author{Jose Pe$\tilde{\mathrm{n}}$a Garcia}
\affiliation{Univ. Grenoble Alpes, CNRS, Institut N\'eel, 38042 Grenoble, France}
\author{Vincent Jeudy}
\affiliation{Laboratoire de Physique des Solides, Universit\'{e} Paris-Saclay, CNRS, 91405 Orsay, France}

\author{André Thiaville}
\affiliation{Laboratoire de Physique des Solides, Universit\'{e} Paris-Saclay, CNRS, 91405 Orsay, France}
\author{Jan Vogel}
\affiliation{Univ. Grenoble Alpes, CNRS, Institut N\'eel, 38042 Grenoble, France}

\author{Marlio Bonfim}
\affiliation{Dep. de Engenharia Elétrica,
Universidade Federal do Parana, Curitiba, Brasil}
\author{Laurent Ranno}
\affiliation{Univ. Grenoble Alpes, CNRS, Institut N\'eel, 38042 Grenoble, France}
\author{Dafine Ravelosona}
\affiliation{Spin-Ion Technologies; 10 Boulevard Thomas Gobert, 91120 Palaiseau, France  }
\affiliation{Centre de Nanosciences et de Nanotechnologies, CNRS, Université Paris-Saclay, 10 boulevard Thomas Gobert, 91120 Palaiseau, France}
\author{Stefania Pizzini}
\affiliation{Univ. Grenoble Alpes, CNRS, Institut N\'eel, 38042 Grenoble, France}
\email{stefania.pizzini@neel.cnrs.fr}

\begin{abstract}
Magnetization reversal and domain wall dynamics in Pt/Co/AlOx trilayers have been tuned
by He$^+$ ion irradiation. Fluences up to 1.5$\times$10$^{15}$ ions/cm$^2$ strongly decrease the perpendicular magnetic anisotropy
(PMA), without affecting neither the spontaneous magnetization nor the strength of the Dzyaloshinskii-Moriya interaction (DMI). This allows us to confirm the
robustness of the DMI interaction against interfacial chemical intermixing, already predicted by theory. In parallel with he decrease of the PMA in the irradiated samples, a strong decrease of the depinning field is observed. This allows the domain walls to reach large maximum velocities with lower magnetic fields with respect to those needed for the pristine
films. Decoupling PMA from DMI can therefore be beneficial for the design of low energy devices based on domain wall dynamics. When the samples are irradiated with larger He$^+$ fluences, the magnetization gets close to the out-of-plane/in-plane reorientation transition
where $\approx$100nm size magnetic skyrmions are stabilized. We observe that as the He$^+$ fluence increases, the skyrmion size decreases while these
magnetic textures become more stable against the application of an external magnetic field.

\end{abstract}

\maketitle


 The study of ultrathin ferromagnetic layers
deposited on heavy atoms in  non-centrosymmetric stacks has received much attention since the discovery that the interfacial Dzyaloshinskii-Moriya interaction (DMI) \cite{Dzyaloshinskii1957,Moriya1960,Levy1980} can lead to the stabilization of non trivial magnetic textures such as chiral N\'{e}el walls and skyrmions \cite{Thiaville2012,Fert2013}. The presence of DMI strongly affects  the  domain wall dynamics, reducing the effect of the Walker breakdown that generally limits their velocity for high magnetic  fields \cite{Thiaville2012,Pham2016}.   One of the most critical technological issues that hinders the application of domain walls or skyrmions to performing spintronic devices, is their interaction with defects. Defects can pin these magnetic textures, limiting their velocities for small driving torques and preventing reproducible displacement events. 

Ion irradiation with  light ions is an appealing way to finely tune the magnetic properties of thin magnetic films and in particular the perpendicular magnetic anisotropy \cite{Chappert1998,Fassbender2004}. Early studies on Pt/Co/Pt multilayers showed that He$^+$ ions with energy in the 30 keV range provoke  short-range (0.2-0.5 nm.) atomic displacements through low energy collisions and implant into the substrate \cite{Chappert1998,Ferre1999,Devolder2000,Devolder_epjb_2001}.   The resulting intermixing at the Co/Pt interfaces  evolves with increasing fluences, therefore decreasing the interfacial perpendicular magnetic anisotropy (PMA) which is known to depend on the anisotropy of the chemical environment at the interface \cite{Weller1995}. 

Since it leads to a modulation of domain wall energy and domain wall width, tuning the PMA is then expected to affect the domain wall dynamics both in the creep and in the flow regime \cite{HerreraDiez2019,Zhao2019}.  
On the other hand, the modification of the DMI strength by ion irradiation \cite{Balk2017, HerreraDiez2019,Zhao2019,Sud2021,deJong2022,Juge2021} has also shown to have an impact on skyrmion stability and dynamics  \cite{Juge2021} and is expected to affect the velocity of chiral Néel walls \cite{Pham2016,Krizakova2019}.


In this work we present the effect of He$^+$ irradiation on the dynamics of chiral domain walls in Pt/Co/AlOx trilayers, a model system with large DMI and out-of-plane magnetization.  We show that the interfacial PMA decreases as the He$^+$ fluence  increases, leading to a progressive decrease of the depinning field. For fluences of the order of 1.5$\times$10$^{15}$ ions/cm$^2$,  domain wall velocities up to 250 m/s  can then be reached for fields two times smaller than in the as-deposited (pristine) sample. Remarkably, the strong decrease of the PMA arising from increased interdiffusion is not accompanied by a decrease of the DMI constant,   allowing us to confirm the robustness of the DMI interaction against intermixing, already predicted by theory \cite{Zimmermann2018}. Decoupling PMA from DMI can be beneficial for the implementation of devices based on chiral domain walls, as their pinning by defects can be considerably reduced while conserving the large maximum velocities. 
\\

 

A Ta(4)/Pt(4)/Co(1.1)/Al(2) magnetic stack (thicknesses in nm) was deposited by magnetron sputtering on Si/SiO$_2$ wafers and the Al layer was oxidised with an oxygen plasma. The sample was diced into small pieces; one of them was kept in the pristine state, while the others were irradiated  with fluences  F1=2$\times$10$^{14}$,  F2=4$\times$10$^{14}$, F3=6$\times$10$^{14}$, F4=1$\times$10$^{15}$, F5=1.5$\times$10$^{15}$, F6=2$\times$10$^{15}$ and F7=3$\times$10$^{15}$ He$^+$/cm$^{2}$ at room temperature. 

The magnetic stacks exhibit out-of-plane magnetization both before and after irradiation. The spontaneous magnetization $M_s$ and the in-plane saturation field $\mu_0H_K$ for the pristine sample and those irradiated with fluences F2, F4, F5 and F7 were measured by Superconducting Quantum Interference Vibrating Sample Magnetometry (VSM-SQUID) and the results are shown in Table I. 


Field-driven domain wall dynamics  was measured by polar magneto-optical Kerr microscopy.  30 ns long pulses up to $B_z$=250 mT  were delivered using a 200~$\mu$m-diameter microcoil associated to a fast pulse current generator.  The
film magnetization was first saturated in the out-of-plane direction. An opposite  magnetic field pulse  was then applied to nucleate a reverse domain. The domain wall (DW) velocity was obtained as the ratio of the displacement of the domain walls during the application of the magnetic field pulses and the total duration of the pulses.  
In order to obtain the strength of the $H_{DMI}$ field, from which the DMI constant $D$ and the sign of the DW chirality can be extracted, domain wall velocities were also measured  as a function of a static in-plane magnetic field, using $B_z$=110~mT as driving field in the flow regime. The DMI of the pristine sample and that for fluence F7 was also measured by Brillouin light scattering (BLS), as detailed in the Supporting Information file.

 \begin{figure}[b]
    \centering
    \includegraphics[scale=0.7]{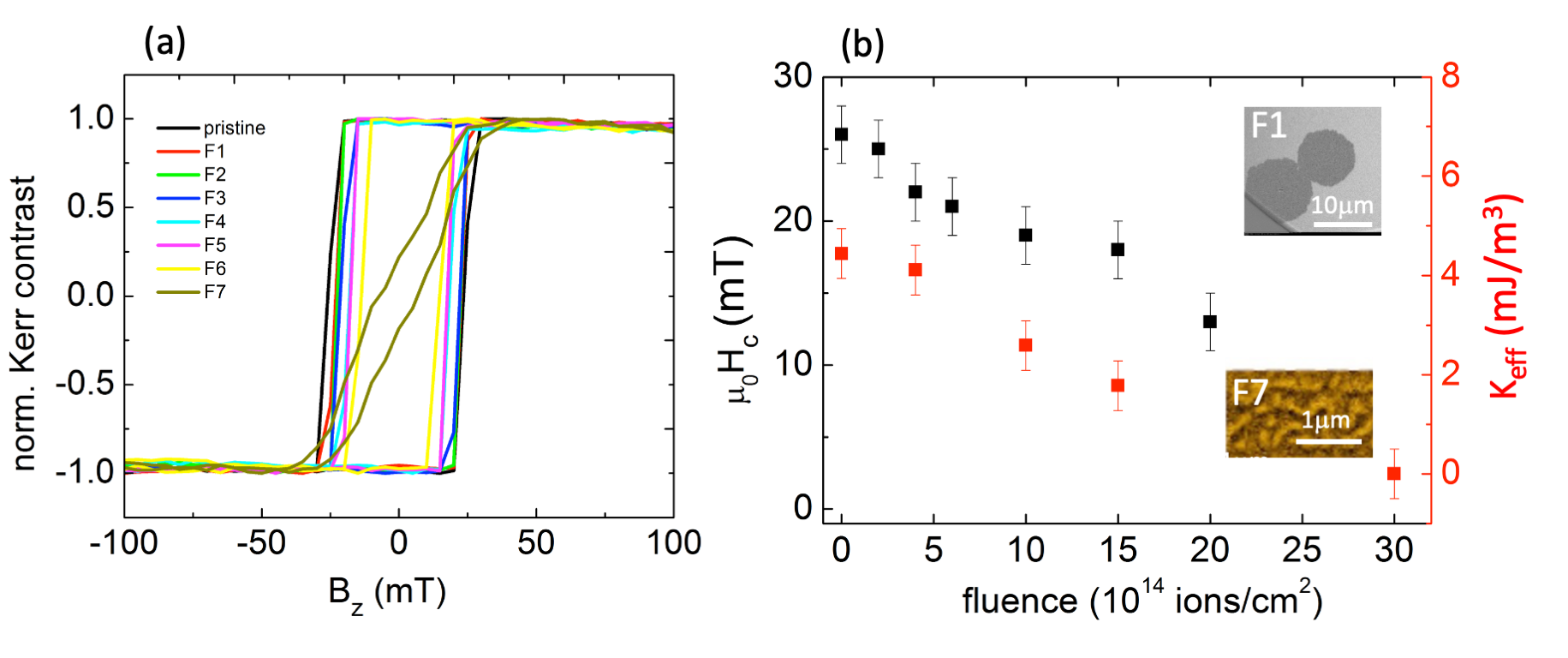}
    \caption{\textbf{Variation of the coercive field of the  Pt/Co/AlOx sample versus He$^+$ ion irradiation }
    (a) Hysteresis loops measured for the Pt/Co/AlOx irradiated with fluences F1 to F7;   (b) Coercive field $\mu_0H_c$  and effective anisotropy energy $K_{eff}$ versus irradiation fluence; (inset): Magnetic domains in the pristine state (top, Kerr microscopy) and after irradiation with fluence F7 (bottom, MFM).}
    \label{fig:1}
\end{figure}

\begin{table*}[b]
\caption[]{\textbf{Micromagnetic parameters measured for the Pt/Co/AlOx samples in the pristine state and after irradiation with He$^{+}$ ions.} Spontaneous magnetization $M_{s}$, in-plane saturation field $\mu_{0}H_{K}$, effective anisotropy energy $K_{eff}$, domain wall energy  $\sigma=4\sqrt{AK_{eff}}-\pi D$, saturation domain wall velocity or Walker velocity  $v_{sat}$, DMI field $\mu_{0}H_{DMI}$, interface DMI constant $D_s$=$Dt$ where $D$ is the DMI constant (in mJ/m$^2$) and $t$ is the Co layer thickness (1.1 nm); $D_{s}^{H}$ is extracted from the measured DMI field; $D_{s}^{W}$ is extracted from $v_{sat}$; $D_{s}^{BLS}$ is obtained with BLS measurements.     The domain wall energy was calculated using $D_{s}^{W}$ and $A$=18 pJ/m; the latter was also used to obtain $D_{s}^{H}$ from the $H_{DMI}$ field. Note that the error bars  indicated  for the pristine sample  are of the same order of magnitude for the irradiated samples.  }  

\begin{tabular}{llllllllllll} \\
 Sample & He$^{+}$ fluence & $M_{s}$ &$\mu_{0}H_{K}$& $K_{eff}$  &  $v_{sat}$ &   $\mu_{0}H_{DMI}$ &    $D_{s}^{H}$ & $D_{s}^{W}$ & $D_{s}^{BLS}$ & $\sigma$  \\
                   &   [ions/cm$^{2}$] & [MA/m ]  &    [mT]  &     [10$^{5}$J/m$^3$]   &   [m/s]   &  [mT]         & [pJ/m]   &   [pJ/m] & 
                   [pJ/m] & [mJ/m$^2$]\\ \\ \hline\\
 \\
pristine & & 1.13$\pm$0.05 & 785 & 4.44  & 260$\pm$20 &   130$\pm$10  &   1.03 $ \pm$0.15 &   1.17$\pm$0.1 & 1.19 $ \pm$0.06 & 7.96 \\
F2  & 4$\times$10$^{14}$ & 1.17 & 730 & 4.26  & 260 &   130  &   1.07 &   1.17  & & 7.62   \\
F4  & 1$\times$10$^{15}$ & 1.16 & 445 & 2.59  &  260 &   100  &  1.07 & 1.20 & & 5.21\\
F5  & 1.5$\times$10$^{15}$ & 1.17 & 305 & 1.78  & 260 &       &        & 1.21  & & 3.71\\
F7  & 3$\times$10$^{15}$ & 1.08   &     &     &     &      &            &  & 0.9  $ \pm$0.05 &
\\
\\

\hline

    \end{tabular}
\end{table*}

The VSM-SQUID measurements show that the spontaneous magnetization is, within the error bars evaluated to be around 5\%,  unchanged up to the irradiation with fluence F5, while a  slight decrease of $M_s$ is measured for the highest fluence F7.  The in-plane saturation field decreases as the fluence increases.  The  effective anisotropy energy $K_{eff}$=1/2$\mu_0H_kM_s$ changes from 4.44$\times$10$^{5}$ J/m$^3$ in the pristine sample to 1.78$\times$10$^{5}$ J/m$^3$ for fluence F5 (see Table I). 

The change of magnetic anisotropy is reflected also in the shape of the hysteresis loops (Figure \ref{fig:1}(a,b)): up to fluence F6 the hysteresis loops have a square shape, indicating that the reversal is dominated by domain wall propagation, with a decrease of the coercivity as the fluence increases. For fluence F7, the measured  butterfly loop suggests the presence of a demagnetized state with labyrinthine domains, indicating that the sample is now close to the reorientation transition from out-of-plane (OOP) to in-plane (IP) magnetization. This is confirmed by the magnetic force microscopy (MFM) measurements that are discussed below.

\begin{figure}[t]
    \centering
    \includegraphics[scale=0.6]{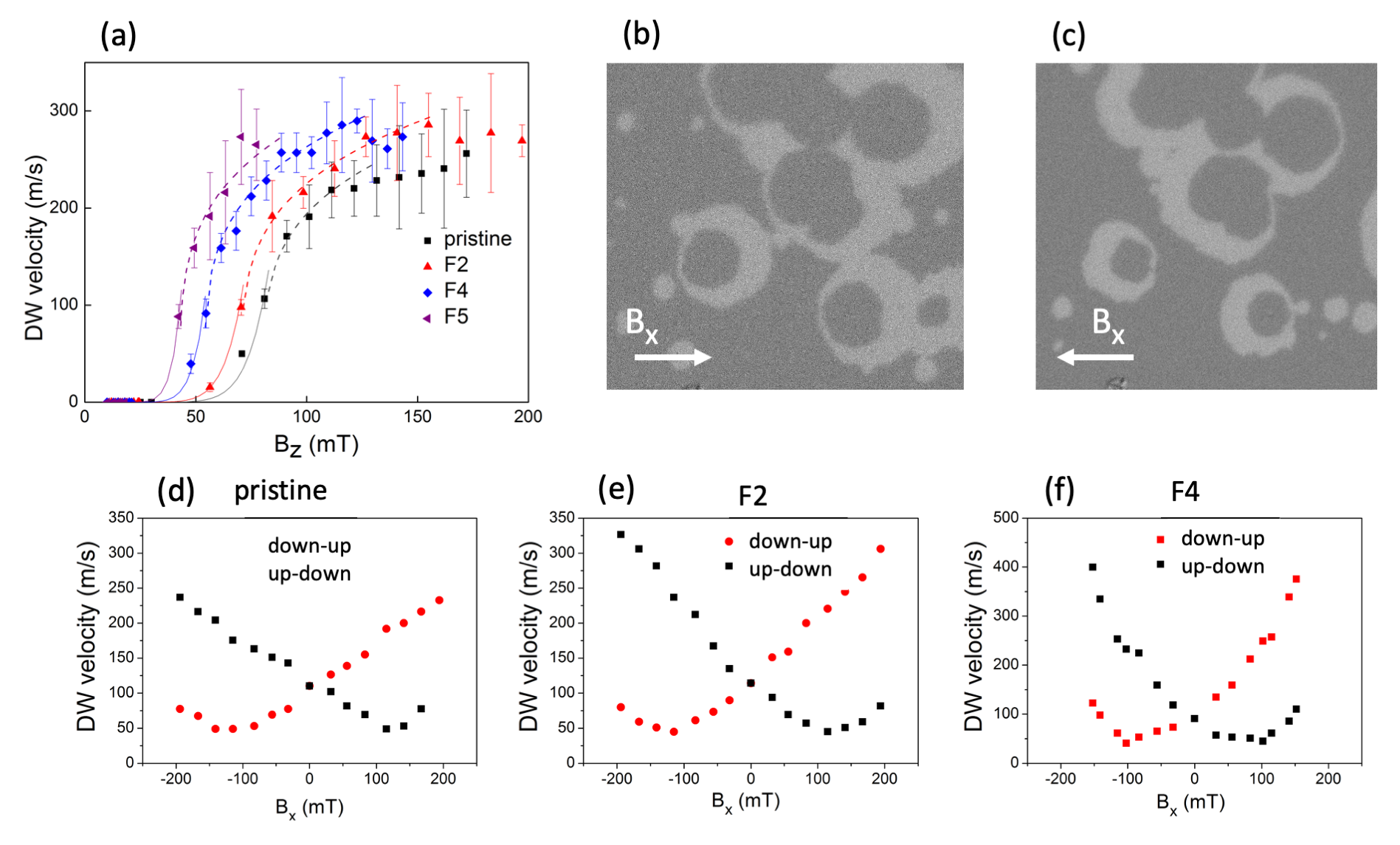}
    \caption{\textbf{Effect of He$^+$ ion irradiation on the dynamics of domain walls in the flow regime}:
    (a) domain wall velocity vs. $B_z$ for the pristine sample and after irradiation with fluences  F2, F4 and F5; (b-c) Differential Kerr images after irradiation with fluence F5, showing the propagation of domain walls driven by $B_z$ in the presence of a constant in-plane  field $B_x$ in opposite directions; (d-f) Domain wall velocity vs. $B_x$ field for down-up and up-down domain walls.   }
    \label{fig:2}
\end{figure}

Figure \ref{fig:2}(a) presents the DW velocity vs. $B_z$ curves measured for the sample in the pristine state and after irradiation with fluences F2, F4 and F5.  Two main trends can  be underlined. First, as the irradiation fluence increases, the depinning field $H_{dep}$ strongly decreases: it is situated at around 80 mT for the pristine sample, and decreases down to 35 mT for fluence F5.  

Secondly, the curves present the features expected for a sample with large DMI : large domain wall velocities (up to $\approx$ 260 m/s),  absence of Walker breakdown after the Walker field and saturation  of domain wall velocity at high fields \cite{Thiaville2012,Pham2016,Krizakova2019,PenaGarcia2021}. Remarkably, within the precision of the measurement, the saturation velocity v$_{sat}$  does not depend on the irradiation fluence. Since $v_{sat} \approx\gamma \pi D/(2 M_s)$ \cite{Pham2016,Krizakova2019} and the magnetization is unchanged, this indicates that, up to fluence F5,  the DMI constant $D$ is not affected by the irradiation. This is confirmed by the DMI values reported in Table I, obtained  using the  experimental $M_s$ in the expression of $v_{sat}$.  Within the error bars estimated at around 10\%, the interface DMI constant $D_s$=$Dt$ (with $t$ the Co layer thickness)  is of $\approx$ 1.2 pJ/m before and after irradiation. A similar value is confirmed by BLS measurement for the pristine sample.

Figure \ref{fig:2}(c-e) present the DW velocity versus in-plane magnetic field $B_x$ measured for the pristine sample and the samples irradiated with fluences F2 and F4. The asymmetric nature of the curves underlines that, in all the samples the DMI stabilizes anticlockwise spin
rotation within the DWs (negative $D$ values). The  DW velocity curves reach a minimum when $B_x$
compensates the $H_{DMI}$ field that stabilizes the DWs in the Néel configuration. 
The DMI constants  can then be obtained from the expression $\mu_0 H_{DMI}=D/(M_{s}\Delta)$, where $\Delta=\sqrt{A/K_{eff}}$ is the DW parameter \cite{Je2013}. 
Using $A$=18 pJ/m in the above expression we obtain DMI constants slightly smaller than those derived from the saturation domain wall velocity but, within the precision of the experiments, we confirm that they are insensitive to irradiation. 

Let us summarise the results described up to now:   irradiation with He$^+$ fluences up to F5=1.5 x 10$^{15}$ ions/cm$^2$ leads to a decrease of the magnetic anisotropy energy causing a decrease of the DW depinning field, with no decrease of the DMI constant. As a consequence, the large domain wall velocities observed  in the flow regime for $B_z$=100 mT in the pristine sample can be obtained  for much lower magnetic fields after irradiation. 

The reduction in anisotropy is in line with the results obtained in the early works of Devolder et al. \cite{Devolder2000,Devolder2001,Devolder_epjb_2001} on Pt/Co/Pt trilayers, who attributed this effect to the soft intermixing of the Pt/Co interfaces induced by short-range atomic displacements. Such intermixing leads to the decrease of the anisotropy of the Co environment which, via the spin-orbit interaction, is at the origin of the PMA \cite{Weller1995}. Since in Pt/Co/AlOx the main source of PMA is at the Pt/Co interface, the same arguments can be applied to our system.

In this respect, it may be surprising that the strong decrease of the interfacial PMA due to chemical mixing is not accompanied by a decrease of the DMI, as it may be expected from the interfacial origin of this parameter. However our results are in line with those obtained for  Pt/Co/Pt trilayers by Lavrijsen et al. \cite{Lavrijsen2015}, who tuned the quality of the Pt/Co top interface by adjusting the Ar pressure  in the magnetron sputtering chamber. Lower pressures produced  more interdiffused interfaces, with a reduction of the  PMA by a factor 3, while keeping almost constant both the magnetization and the DMI field. 

The robustness of DMI against intermixing found in this work has been confirmed by the results of density functional theory of Zimmermann et al. \cite{Zimmermann2018} for Pt/Co bilayers.  Starting from a perfect interface, the calculations predict that  intermixing 20\% of the interface atoms causes  a slight decrease of DMI, after which the DMI value can be kept constant until more than 80\% of the atoms are displaced. This behaviour is obtained when the contribution to the DMI of Co-Co pairs  away from the initial interface are taken into account, while a strong drop of the DMI is obtained if only the Co-Co pairs situated along the initial  interface are considered.

Up to now we have only considered the effect of the irradiation on the Pt/Co interface.  However, since in the Pt/Co/AlOx stack both Pt/Co and Co/oxide interfaces give a contribution to both PMA \cite{Monso2002,Manchon2008a,Manchon2008b,Manchon2008c} and DMI \cite{Chaves2019} we cannot a~priori exclude the impact of the irradiation on the Co/oxide interface, as was done in previous studies \cite{Juge2021}. 

\begin{figure}[t]
    \centering
    \includegraphics[scale=0.8]{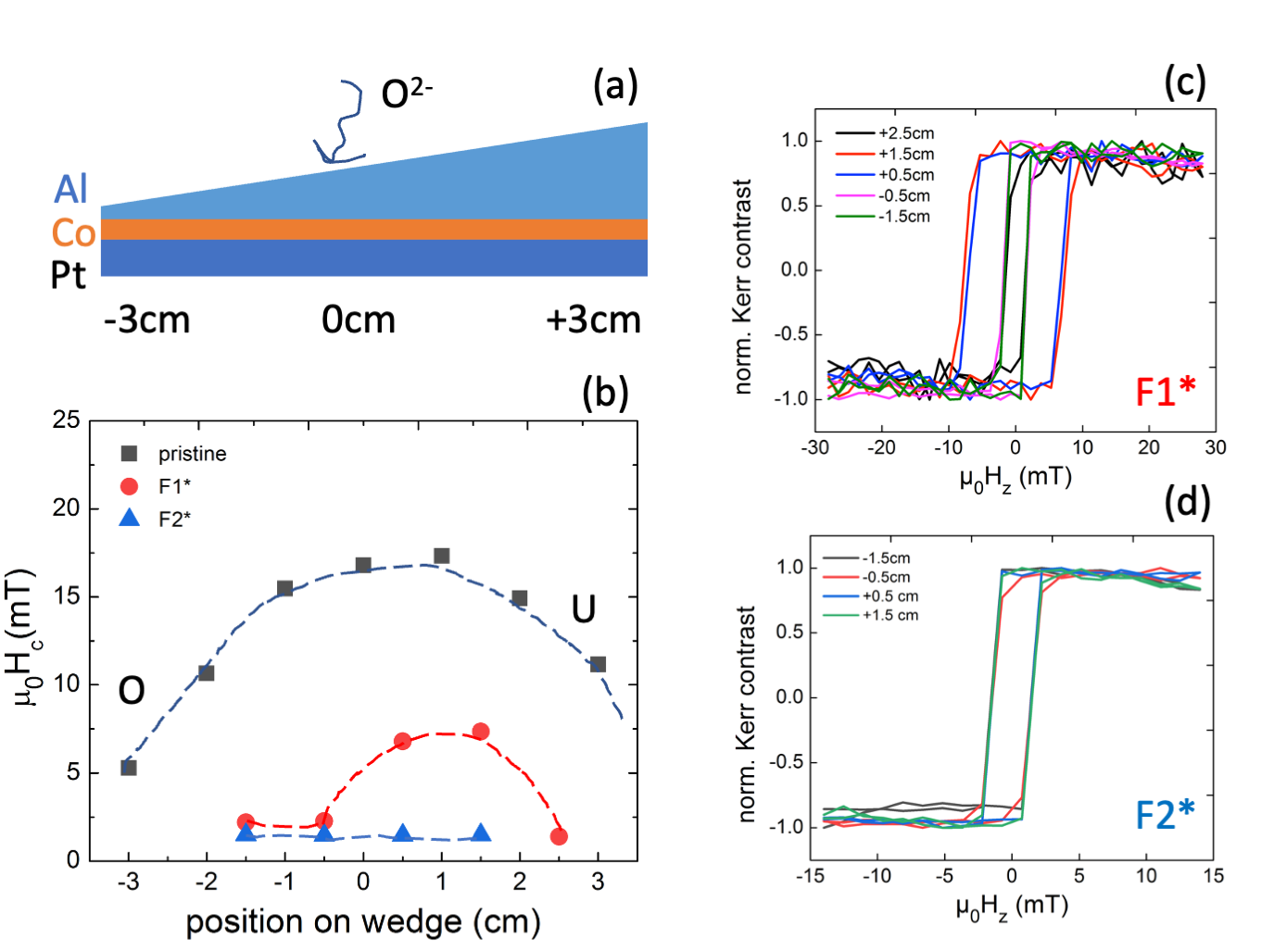}
    \caption{\textbf{Effect of the He$^+$ irradiation on the Co/AlOx interface anisotropy} (a) Sketch of the Pt/Co/AlOx sample with wedge of Al exposed to an oxygen plasma; (b) Coercive field vs. sample position i.e. versus oxidation rate in the Pt/Co/AlOx sample  in the pristine state and after irradiation with fluence F1*=1.5$\times$10$^{14}$ and F2*=4$\times$10$^{14}$ ions/cm$^2$; U and O stand for under-oxidized and over-oxidized Co layer;  (c-d) hysteresis loops vs. sample position measured for fluences F1* and F2*.  
    }
    \label{fig:3}
\end{figure}

To clarify this issue, we have grown a Ta(3)/Pt(4)/Pt/Co(1.1)/Al trilayer in which the Al layer is a wedge with varying thickness.  The exposure to an oxygen plasma  gives rise to a gradient of Co oxidation and as a consequence to a variation of PMA along the wedge \cite{Monso2002,Manchon2008a,Manchon2008b}. The variation of the coercive field  along the wedge is shown in Figure \ref{fig:3}  in the pristine state and after irradiation with fluence F1*=1.5$\times$10$^{14}$ and F2*=4$\times$10$^{14}$ ions/cm$^2$. The parabolic curve obtained for the pristine sample reflects the well known dependence of the PMA of the Co/oxide interface on its oxidation degree \cite{Monso2002,Manchon2008a,Manchon2008b}. The maximum of the curve (around position +0.5cm on the wedge in Figure \ref{fig:3}) corresponds to the situation where the interfacial Co atoms are all bonded to oxygen atoms \cite{Manchon2008a}.  After irradiation with the lowest fluence F1*=1.5 $\times$ 10$^{14}$ ions/cm$^2$ the coercive field (and therefore the PMA \cite{Givord2003}) decreases, and the maximum of the coercive field  is displaced towards  sample position +1cm, corresponding to larger Al thickness, while in the positions corresponding to  thinner Al the magnetization turns in-plane.  The direction of the displacement of the coercivity curve indicates that the degree of oxidation of the Co/AlOx interface increases, probably due to transport of oxygen ions towards this interface, as indicated also by the Transport of Ions in Matter (TRIM) code calculations \cite{Ziegler2010} (not shown here).  When the fluence is increased to F2*=4$\times$10$^{14}$ ions/cm$^2$ the coercivity  versus sample position curve becomes flat: this suggests that after this irradiation
the Co/AlOx interface composition becomes homogeneous, independently on the thickness of the AlOx layer. The exact mechanism occurring at the interface is difficult to capture. We tend to exclude the possibility that, following the intermixing between Al and Co,  the Co/AlOx interface no longer contributes to the PMA : without this contribution  the magnetization of  Pt/Co/AlOx would be in-plane i.e. the anisotropy of the Pt/Co interface would not be sufficient to stabilize an out-of-plane magnetization. A possible explanation of this effect is that the He$^+$ irradiation together with the large oxygen affinity of aluminium trigger the evolution of the Al layer towards a thermodynamically more stable state,  corresponding with it  being more homogeneously oxidized. This could explain the loss of the dependence of PMA on sample position.

\begin{figure}[t]
    \centering
    \includegraphics[scale=0.5]{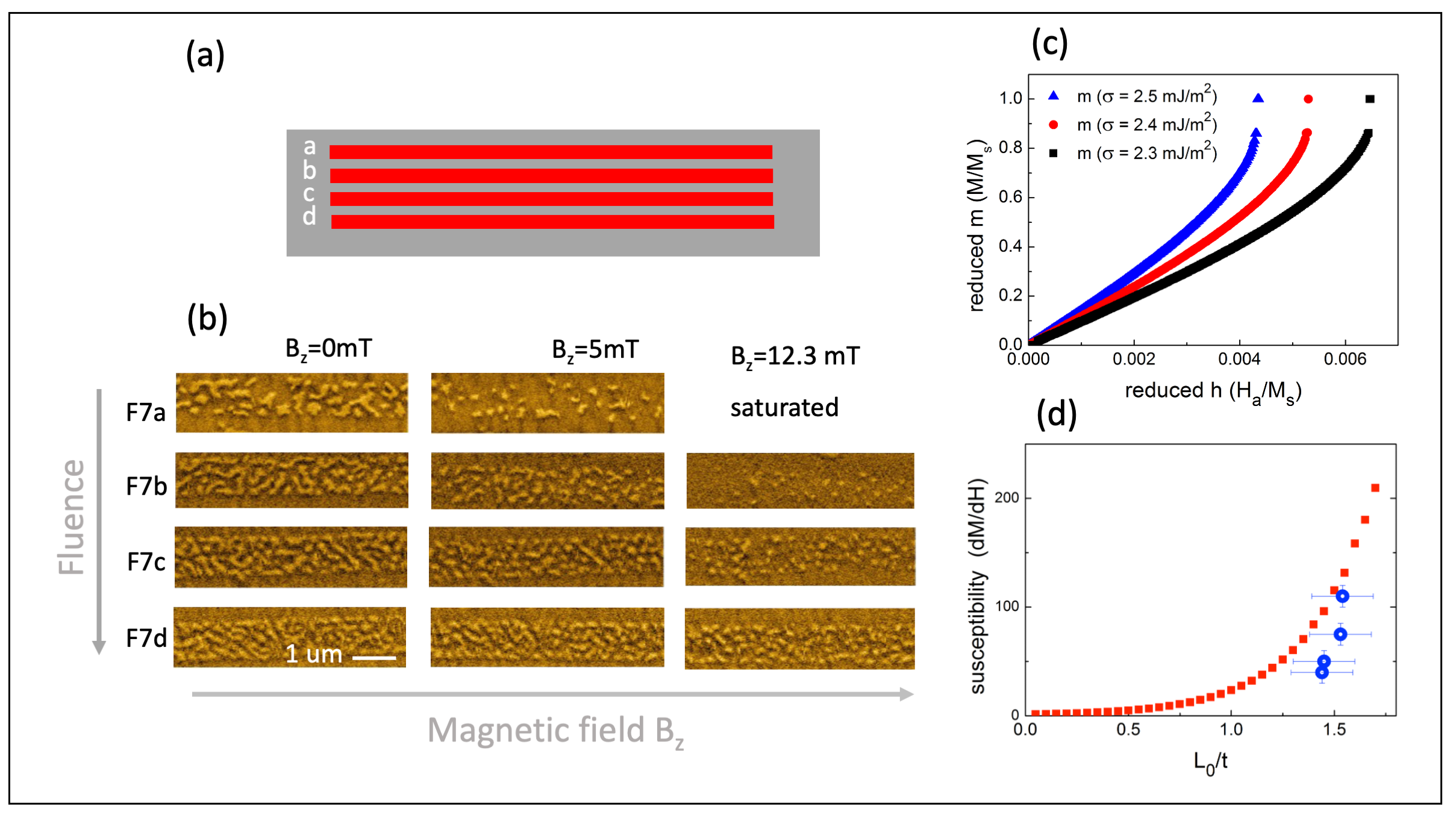}
    \caption{\textbf{Stripe domains and skyrmions after irradiation with large fluence stabilizing stripe domains and skyrmions}: (a) sketch of the sample with the 1 $\mu$m-wide irradiated areas : F7a= 2.5 $\times$10$^{15}$ ions/cm$^2$), F7b= 2.7$\times$ 10$^{15}$ ions/cm$^2$, F7c= 2.9 $\times$10$^{15}$ ions/cm$^2$, F7d= 3.2 $\times$10$^{15}$ ions/cm$^2$) (b) MFM images showing the presence of stripe domains for B$_z$=0, that are gradually transformed into skyrmion bubbles with an applied B$_z$ field;  (c) magnetic susceptibility $\chi$=$M_s/H_z$ as a function of reduced  dipolar length $L_{0}/t$, calculated from Ref. \cite{Kooy1960}; $M_s$=1.08 MA/m and $t$=1.1 nm were used in the calculation of $L_0$. The blue dots are the experimental susceptibility values $\chi^{exp}=M_s/B_a$.  
    }
    \label{fig:4}
\end{figure}

We now discuss the magnetic textures observed in the Pt/Co/AlOx stack (with fixed Al thickness) irradiated with larger fluences, around F7$\approx$3$\times$10$^{15}$ ions/cm$^2$. The butterfly-like hysteresis loop (Figure 1) indicates that  the PMA  is strongly reduced and the magnetization easy axis is close to the out-of-plane/in-plane reorientation transition. Also, the measurements carried out by BLS indicate that the interfacial DMI has strongly decreased. To better follow the evolution of the magnetic properties, the sample was irradiated locally with a focused  He$^+$ beam, into 1$\mu$m  $\times$ 50 $\mu$m stripes, using fluences F7a = 2.5 $\times$ 10$^{15}$ ions/cm$^2$, F7b = 2.7 $\times$ 10$^{15}$ ions/cm$^2$, F7c= 2.9 $\times$ 10$^{15}$ ions/cm$^2$ and  F7d= 3.2 $\times$ 10$^{15}$ ions/cm$^2$ (see sketch in Figure \ref{fig:4}(a)). The magnetic textures stabilized in these zones, imaged by MFM,  are shown in Figure \ref{fig:4}(b). 
They consist of stripe domains and skyrmion bubbles whose average dimensions decrease from $\approx$ 140nm to $\approx$ 100nm as the fluence increases (Table II). 

In the case of ultrathin ferromagnetic layers, the labyrinthine domain width $L$ (and the corresponding skyrmion size)  is given by  $L=C~t~\exp(\pi L_{0}/t)$, where $L_{0}=\sigma/\mu_{0}M_{s}^{2}$ is the characteristic dipolar length, $\sigma=4\sqrt{AK_{eff}}-\pi D$ is the domain wall energy, $t$ is the ferromagnetic film thickness and $C$ is a numerical constant of the order of 1 \cite{Schafer1998,Kaplan1993}. 
Using the value of $M_s$  measured for fluence F7 ($M_s$=1.08 MA/m) and $t$=$t_{Co}$=1.1 nm in the expression of $L$, we obtain DW energy values decreasing from 2.48 mJ/m$^2$ to  2.33 mJ/m$^2$ for fluences  F7a to F7d (see Table II). This reflects  the expected decrease of the effective anisotropy as the fluence increases. Note that our approach is validated by the fact that these DW energy values are lower than those obtained for fluence F5 using the measured magnetic parameters. 


\begin{table*}[t]
\caption[]{Average stripe domain or skyrmion size $L$ measured from  MFM images after irradiation with fluences F7a to F7d, experimental annihilation field $B_a^{exp}$ and  susceptibility $\chi^{exp}=\mu_0M_s/B_a$,  reduced dipolar length $L_0/t$ and domain energy $\sigma$ calculated from $L=Ct~\exp(\pi L_{0}/t)$, with $L_{0}=\sigma/\mu_{0}M_{s}^{2}$. Note that the error bars  indicated  for fluence F7a  are of the same order of magnitude for all the fluences.  }
\begin{tabular}{p{12mm}p{25mm}p{18mm}p{18mm}p{15mm}p{15mm}p{15mm}}
\hline
 \\
  & He$^{+}$ fluence & $L$ & $B_a^{exp}$ & $\chi^{exp}$ & $L_{0}/t$ & $\sigma$ \\
 &   [ions/cm$^{2}$] &  [nm] &   [mT]  & & &  [mJ/m$^2$]   \\ 
                   \hline
                  \hline
 \\
F7a & 2.5$\times$110$^{15}$    & 138 $\pm$ 5 &  12 $\pm$ 3 & 110 $\pm$ 10 &1.54    &  2.48    \\
F7b  &  2.7$\times$110$^{15}$   & 136        &   18 & 75  &         1.53 &     2.47 \\
F7c  & 2.9$\times$110$^{15}$ &   105         &  27 & 50   &         1.45  &    2.34  \\
F7d  & 3.2$\times$110$^{15}$ &   103         &  35 &  40  &           1.44 &     2.33  \\

\hline

    \end{tabular}
\end{table*}

A destabilizing  out-of-plane magnetic field, applied in the direction of the skyrmion core magnetization, leads to the progressive annihilation of the magnetic textures. However the efficiency of this effect decreases as the irradiation fluence increases (Figure \ref{fig:4}(b)). For fluence F7a, the field $B_z$=5 mT strongly decreases the skyrmion density  and $B_z$=12.3 mT  is sufficient to annihilate all the skyrmions. As the fluence is increased, the magnetic skyrmions are much more stable against the application of the magnetic field, indicating that the annihilation field $B_a$ increases as the PMA  decreases (see Table II). 


This behaviour can be explained schematically by taking into account the competition between the interface PMA and the demagnetizing field. In the absence of interface PMA, the magnetic field needed to saturate the magnetization perpendicular to film plane is equal to the demagnetizing field, $\mu_0M_s$. In the presence of interface PMA, this anisotropy helps to pull the magnetization out of plane and the perpendicular saturation field therefore decreases with increasing interface PMA. 

Using the approach developed by Kooy and Enz \cite{Kooy1960} we have calculated numerically the reduced magnetization  curves ($m=M/M_s$) 
 as a function of reduced field ($H_s/M_s)$,   
for three values of the DW energy $\sigma$ in the range of those extracted from the size of the stripe domains after irradiation (Figure \ref{fig:4}(c)).  
The calculation, which assumes periodic magnetic stripes, perpendicular magnetization and  DWs width negligible with respect to the stripe domain period ($\Delta \ll$2$L$), is based on the minimization of the demagnetizing energy, using a given value of $\sigma$. In agreement with the experimental results and the hand-waving arguments, the curves show that the susceptibility decreases and the annihilation field  increase as the DW energy decreases.  



For  clarity, the magnetic susceptibility $\chi$=$M_s/B_z$ was calculated as a function of reduced  dipolar length $L_{0}/t$ (Figure \ref{fig:4}(d)).  We notice that  $\chi$ increases exponentially as $L_0/t$ increases, i.e. as the DW energy increases. Considering the uncertainty of the experimental parameters, our results are in excellent agreement with the calculations. 

In conclusion, we have shown that by modifying the magnetic anisotropy of Pt/Co/AlOx stacks with  He$^+$ irradiation, the dynamics of chiral Néel walls and the size and stability of magnetic skyrmions can be finely controlled.  A strong decrease of the PMA, associated to atomic displacements at both Pt/Co and Co/AlOx  interfaces,  is obtained without deteriorating the strength of the interfacial  Dzyaloshinskii-Moriya interaction. The robustness of the DMI against interfacial intermixing, confirming theoretical  predictions, is beneficial for the dynamics of field-driven domain walls. Indeed, the decrease of the depinning field, associated to the decreasing PMA, together with the persisting large DMI, allow obtaining large DW velocities for much lower magnetic fields in the irradiated samples, compared with the pristine sample. Decoupling the PMA from the DMI can therefore be a route towards the conception of low energy devices based on field or current-driven domain wall dynamics. 

When the He$^+$ fluence is increased to $\approx$ 3 $\times$ 10$^{15}$ ions/cm$^2$ the magnetization easy axis approaches the out-of-plane/in-plane reorientation transition where labyrinthine domains and skyrmions are stabilized. The skyrmion size can be finely tuned with the irradiation fluence and their stability against a magnetic field  is shown to increase exponentially as the irradiation fluence increases. 

Supporting information: Additional sample characterization using Brillouin Light Scattering and description of the methodology used to extract sample parameters.

\section{Acknowledgements}
 We acknowledge the support of the Agence Nationale de la Recherche (project ANR-17-CE24-0025 (TOPSKY) and of the DARPA TEE program through Grant No. MIPR HR0011831554. The authors acknowledge funding from the European Union’s Horizon 2020 research and innovation program under Marie Sklodowska-Curie Grant Agreement No. 754303 and No. 860060 “Magnetism and the effect of Electric Field” (MagnEFi). J.P.G. also thanks the Laboratoire d\textquotesingle Excellence LANEF in Grenoble (ANR-10-LABX-0051) for its support. B. Fernandez, T. Crozes, Ph. David, E. Mossang and E. Wagner are acknowledged for their technical help. We thank Dominique Mailly who carried out the focussed ion beam irradiations.


\begin{thebibliography}{10}

\bibitem{Dzyaloshinskii1957}
I.~E. Dzyaloshinskii.
\newblock Thermodynamical theory of "weak" ferromagnetism in antiferromagnetic
  substances.
\newblock {\em Sov. Phys. JETP}, 5:1259, 1957.

\bibitem{Moriya1960}
T.~Moriya.
\newblock Anisotropic superexchange interaction and weak ferromagnetism.
\newblock {\em Phys. Rev.}, 120:91, 1960.

\bibitem{Levy1980}
A.~Fert and Peter~M. Levy.
\newblock Role of anisotropic exchange interactions in determining the
  properties of spin-glasses.
\newblock {\em Phys. Rev. Lett.}, 44:1538--1541, Jun 1980.

\bibitem{Thiaville2012}
A.~Thiaville, S.~Rohart, E.~Ju{\'{e}}, V.~Cros, and A.~Fert.
\newblock Dynamics of {Dzyaloshinskii} domain walls in ultrathin magnetic
  films.
\newblock {\em EPL}, 100:57002, 2012.

\bibitem{Fert2013}
A.~Fert, V.~Cros, and J.~Sampaio.
\newblock Skyrmions on th track.
\newblock {\em Nature Nanotech.}, 8:152, 2013.

\bibitem{Pham2016}
Thai~Ha Pham, J.~Vogel, J.~Sampaio, M.~Vanatka, J-C Rojas-Sanchez, M.~Bonfim,
  D.~S. Chaves, F.~Choueikani, P.~Ohresser, E.~Otero, A.~Thiaville, and
  S.~Pizzini.
\newblock {Very large domain wall velocities in Pt/Co/GdOx and Pt/Co/Gd
  trilayers with Dzyaloshinskii-Moriya interaction}.
\newblock {\em {EPL}}, {113}:67001, {2016}.

\bibitem{Chappert1998}
C.~Chappert, H.~Bernas, J.~Ferre, V.~Kottler, J.-P. Jamet, Y.~Chen, E.~Cambril,
  T.~Devolder, F.~Rousseaux, V.~Mathet, and H.~Launois.
\newblock Planar patterned magnetic media obtained by ion irradiation.
\newblock {\em Science}, 280:1919, 1998.

\bibitem{Fassbender2004}
J.~Fassbender, D.~Ravelosona, and Y~Samson.
\newblock Tailoring magnetism by light-ion irradiation.
\newblock {\em Journal of Physics D: Applied Physics}, 37:R179, 2004.

\bibitem{Ferre1999}
J.~Ferré, C.~Chappert, H.~Bernas, J.-P. Jamet, P.~Meyer, O.~Kaitasov,
  S.~Lemerle, V.~Mathet, F.~Rousseaux, and H.~Launois.
\newblock Irradiation induced effects on magnetic properties of {Pt/Co/Pt}
  ultrathin films.
\newblock {\em Journal of Magnetism and Magnetic Materials}, 198-199:191, 1999.

\bibitem{Devolder2000}
T.~Devolder.
\newblock Light ion irradiation of {Co/Pt} systems: Structural origin of the
  decrease in magnetic anisotropy.
\newblock {\em Phys. Rev. B}, 62:5794, 2000.

\bibitem{Devolder_epjb_2001}
T.~Devolder, S.~Pizzini, J.~Vogel, H.~Bernas, C.~Chappert, V.~Mathet, and
  M.~Borowski.
\newblock {\em Eur. Phys. J. B}, 22:193, 2001.

\bibitem{Weller1995}
D.~Weller, J.~St\"ohr, R.~Nakajima, A.~Carl, M.~G. Samant, C.~Chappert,
  R.~M\'egy, P.~Beauvillain, P.~Veillet, and G.~A. Held.
\newblock Microscopic origin of magnetic anisotropy in {Au/Co/Au} probed with
  x-ray magnetic circular dichroism.
\newblock {\em Phys. Rev. Lett.}, 75:3752, 1995.

\bibitem{HerreraDiez2019}
L.~Herrera Diez, M.~Voto, A.~Casiraghi, M.~Belmeguenai, Y.~Roussign\'e,
  G.~Durin, A.~Lamperti, R.~Mantovan, V.~Sluka, V.~Jeudy, Y.~T. Liu,
  A.~Stashkevich, S.~M. Ch\'erif, J.~Langer, B.~Ocker, L.~Lopez-Diaz, and
  D.~Ravelosona.
\newblock Enhancement of the {D}zyaloshinskii-{M}oriya interaction and domain
  wall velocity through interface intermixing in {Ta/CoFeB/MgO}.
\newblock {\em Phys. Rev. B}, 99:054431, 2019.

\bibitem{Zhao2019}
X.~Zhao, B.~Zhang, N.~Vernier, X.~Zhang, M.~Sall, T.~Xing, L.~Herrera~Diez,
  C.~Hepburn, G.~Wang, L.~Durin, A.~Casiraghi, M.~Belmeguenai, Y.~Roussigné,
  A.~Stashkevich, S.M. Chérif, J.~Langer, B.~Ocker, S.~Jaiswal, G.~Jakob,
  M.~Kläui, W.~Zhao, and D.~Ravelosona.
\newblock Enhancing domain wall velocity through interface intermixing in
  {W-CoFeB-MgO} films with perpendicular anisotropy.
\newblock {\em Appl. Phys. Lett.}, 115:122404, 2019.

\bibitem{Balk2017}
A.~L. Balk, K-W. Kim, D.~T. Pierce, M.~D. Stiles, J.~Unguris, and S.~M. Stavis.
\newblock Simultaneous control of the {D}zyaloshinskii-{M}oriya interaction and
  magnetic anisotropy in nanomagnetic trilayers.
\newblock {\em Phys. Rev. Lett.}, 119:077205, 2017.

\bibitem{Sud2021}
A.~Sud, S.~Tacchi, D.~Sagkovits, C.~Barton, M.~Sall, L.~H. Diez,
  E.~Stylianidis, N.~Smith, L.~Wright, S.~Zhang, X.~hang, D.~Ravelosona,
  G.~Carlotti, H.~Kurebayashi, O.~Kazakova, and M.~Cubukcu.
\newblock Tailoring interfacial effect in multilayers with
  {D}zyaloshinskii–{M}oriya interaction by helium ion irradiation.
\newblock {\em Scientific Reports}, 11:023626, 2021.

\bibitem{deJong2022}
Mark C.~H. de~Jong, Mari\"elle~J. Meijer, Juriaan Lucassen, Jos van Liempt,
  Henk J.~M. Swagten, Bert Koopmans, and Reinoud Lavrijsen.
\newblock Local control of magnetic interface effects in chiral
  $\mathrm{Ir}|\mathrm{Co}|\mathrm{Pt}$ multilayers using {Ga}$^{+}$ ion
  irradiation.
\newblock {\em Phys. Rev. B}, 105:064429, 2022.

\bibitem{Juge2021}
Roméo Juge, Kaushik Bairagi, Kumari~Gaurav Rana, Jan Vogel, Mamour Sall,
  Dominique Mailly, Van~Tuong Pham, Qiang Zhang, Naveen Sisodia, Michael
  Foerster, Lucia Aballe, Mohamed Belmeguenai, Yves Roussigné, Stéphane
  Auffret, Liliana~D. Buda-Prejbeanu, Gilles Gaudin, Dafiné Ravelosona, and
  Olivier Boulle.
\newblock Helium ions put magnetic skyrmions on the track.
\newblock {\em Nano Letters}, 21:2989, 2021.

\bibitem{Krizakova2019}
V.~Krizakova, J.~Pe\~{n}a Garcia, J.~Vogel, D.~de~Souza~Chaves, S.~Pizzini, and
  A.~Thiaville.
\newblock Study of the velocity plateau of {Dzyaloshinskii} domain walls.
\newblock {\em Phys. Rev. B}, 100:214404, 2019.

\bibitem{Zimmermann2018}
Bernd Zimmermann, William Legrand, Davide Maccariello, Nicolas Reyren, Vincent
  Cros, Stefan Blügel, and Albert Fert.
\newblock {D}zyaloshinskii-{M}oriya interaction at disordered interfaces from
  ab initio theory: Robustness against intermixing and tunability through
  dusting.
\newblock {\em Applied Physics Letters}, 113:232403, 2018.

\bibitem{PenaGarcia2021}
Jose Pe$\tilde{\mathrm{n}}$a~Garcia, Aymen Fassatoui, Marlio Bonfim, Jan Vogel,
  Andr\'e Thiaville, and Stefania Pizzini.
\newblock Magnetic domain wall dynamics in the precessional regime: Influence
  of the {Dzyaloshinskii-Moriya} interaction.
\newblock {\em Phys. Rev. B}, 104:014405, 2021.

\bibitem{Je2013}
S.-G. Je, D-H. Kim, S.-C. Yoo, B.-C. Min, K.-J. Lee, and S.-B. Choe.
\newblock Asymmetric magnetic domain-wall motion by the
  {D}zyaloshinskii-{M}oriya interaction.
\newblock {\em Phys. Rev. B}, 88:214401, 2013.

\bibitem{Devolder2001}
T.~Devolder, S.~Pizzini, J.~Vogel, H.~Bernas, C.~Chappert, V.~Mathet, and
  M.~Borowski.
\newblock X-ray absorption analysis of sputter-grown {Co/Pt} stackings before
  and after helium irradiation.
\newblock {\em The European Physical Journal B - Condensed Matter and Complex
  Systems}, 22:193, 2001.

\bibitem{Lavrijsen2015}
R.~Lavrijsen, D.~M.~F. Hartmann, A.~van~den Brink, Y.~Yin, B.~Barcones, R.~A.
  Duine, M.~A. Verheijen, H.~J.~M. Swagten, and B.~Koopmans.
\newblock Asymmetric magnetic bubble expansion under in-plane field in
  {Pt/Co/Pt}: Effect of interface engineering.
\newblock {\em Phys. Rev. B}, 91:104414, 2015.

\bibitem{Monso2002}
S.~Monso, B.~Rodmacq, S.~Auffret, G.~Casali, F.~Fettar, B.~Gilles, B.~Dieny,
  and P.~Boyer.
\newblock Crossover from in-plane to perpendicular anisotropy in
  {Pt/CoFe/AlO$_x$} sandwiches as a function of al oxidation: A very accurate
  control of the oxidation of tunnel barriers.
\newblock {\em Appl. Phys. Lett.}, 80(22):4157--4159, 2002.

\bibitem{Manchon2008a}
A.~Manchon, C.~Ducruet, L.~Lombard, S.~Auffret, B.~Rodmacq, B.~Dieny,
  S.~Pizzini, J.~Vogel, V.~Uhl\'{\i}\u{r}, M.~Hochstrasser, and G.~Panaccione.
\newblock Analysis of oxygen induced anisotropy crossover in {Pt/Co/MOx}
  trilayers.
\newblock {\em J. Appl. Phys.}, 104:043914, 2008.

\bibitem{Manchon2008b}
A.~Manchon, S.~Pizzini, V.~Uhl\'{\i}\u{r}, J.~Vogel, L.~Lombard, C.~Ducruet,
  S.~Auffret, B.~Rodmacq, B.~Dieny, M.~Hochstrasser, and G.~Panaccione.
\newblock X-ray analysis of oxygen-induced perpendicular magnetic anisotropy in
  {Pt/Co/AlOx} trilayers.
\newblock {\em J. Magn. Magn. Mater.}, 320:1889, 2008.

\bibitem{Manchon2008c}
A.~Manchon, S.~Pizzini, V.~Vogel, J.~Uhl\'{\i}\u{r}, L.~Lombard, C.~Ducruet,
  S.~Auffret, B.~Rodmacq, B.~Dieny, M.~Hochstrasser, and G.~Panaccione.
\newblock X-ray analysis of the magnetic influence of oxygen in
  {Pt/Co/AlO$_{x}$} trilayers.
\newblock {\em J. Appl. Phys.}, 103:07A912, 2008.

\bibitem{Chaves2019}
D.~de~Souza~Chaves, F.~Ajejas, V.~K\v{r}i\v{z}\'{a}kov\'{a}, J.~Vogel, and
  S.~Pizzini.
\newblock Oxidation dependence of the {Dzyaloshinskii-Moriya} interaction in
  {Pt/Co/MO$_{x}$} trilayers ( {M = Al or Gd}).
\newblock {\em Phys. Rev. B}, 99:144404, 2019.

\bibitem{Givord2003}
Dominique Givord, Michel Rossignol, and Vitoria~M.T.S. Barthem.
\newblock The physics of coercivity.
\newblock {\em Journal of Magnetism and Magnetic Materials}, 258-259:1--5,
  2003.
\newblock Second Moscow International Symposium on Magnetism.

\bibitem{Ziegler2010}
James~F. Ziegler, M.D. Ziegler, and J.P. Biersack.
\newblock Srim – the stopping and range of ions in matter (2010).
\newblock {\em Nuclear Instruments and Methods in Physics Research Section B:
  Beam Interactions with Materials and Atoms}, 268:1818--1823, 2010.
\newblock 19th International Conference on Ion Beam Analysis.

\bibitem{Kooy1960}
C.~Kooy and U.~Enz.
\newblock Experimental and theoretical study of the domain configuration in
  thin layers of {BaFe$_{12}$O$_{19}$}.
\newblock {\em Philips Res. Rep.}, 15:7, 1960.

\bibitem{Schafer1998}
A.~Hubert and R.~Sch{\"a}fer.
\newblock {\em Magnetic Domains, The Analysis of Magnetic Microstructures}.
\newblock Springer: Berlin, 1998.

\bibitem{Kaplan1993}
B.~Kaplan and G.~A. Gehring.
\newblock The domain structure in ultrathin magnetic films.
\newblock {\em J. Magn. Magn. Mater.}, 128:111, 1993.

\end{thebibliography}

\end{document}